\documentstyle[aps,pre]{revtex}

\begin{document}
\draft
\title{On the validity of the Boltzmann equation to describe low density
granular systems}
\author{J. Javier Brey and  M.J. Ruiz--Montero}
\address{F\'{\i}sica Te\'{o}rica, Universidad de Sevilla, Apdo.\ de
 Correos 1065, E-41080 Sevilla, Spain}
\date{today}

\maketitle

\begin{abstract}

The departure of a granular gas in the instable region of parameters
from the initial homogeneous cooling state is studied. Results from
Molecular Dynamics and from Direct Monte Carlo simulation of the Boltzmann
equation are compared. It is shown that the Boltzmann equation accurately
predicts the low density limit of the system. The relevant role played by the
parallelization of the velocities as time proceeds and the dependence of 
this effect on the density is analyzed in detail.

\end{abstract}

\section{Introduction}
\label{s1}

The study of granular systems has attracted much interest in the 
last years. The practical importance of these systems, which are present
in many situations in daily life, together with the richness and
complexity of their behavior\cite{JNyB96}, have motivated  many researches 
trying to describe and understand the physical  mechanisms governing granular 
flows. Of course, the variety of what we call ''granular media'' 
makes it necessary to use different theoretical descriptions
depending on the problem we wish to address. In the context
of rapid, low density, granular flows, the application of the methods of
the kinetic theory for molecular (elastic) systems has proven
to be a very useful tool. The starting point for this description
has been in many cases the extension to inelastic collisions of the
Boltzmann equation, which is derived under the same hypothesis
as in elastic systems. From this kinetic equation, closed hydrodynamic
equations with explicit expression for the fluxes and the associated transport 
coefficients (up to second order in the gradients) for inelastic hard 
particles have been derived \cite{BDKyS98,ByC00}. This hydrodynamic  
picture  has been found to provide an accurate 
description of granular systems in very different situations, driven and not
driven. The  Boltzmann equation has also been used to
study  the velocity distribution of a granular gas, modeled
in most of the cases as a system of inelastic hard particles. 
Of course, the shape of the distribution depends on the state of the
system. The simplest possibility is the so-called homogeneous cooling 
state (HCS), the state of a homogeneous, freely evolving  granular gas 
\cite{C90}. In that case, the Boltzmann equation is shown to admit a 
solution whose  time dependence can be scaled through its second  moment. 
Deviations from gaussianity in the HCS have been quantified  by computing 
the kurtosis of  the distribution \cite{GyS95,NyE98,BRyC96} and by  
establishing the  existence of exponential velocity 
tails \cite{NyE98,EyP97,BCyR99}. The possible solutions of the Boltzmann
equation in the case of a vibrated system in absence of gravity have also 
been investigated, and  the conditions  for the existence of a steady 
solution whose space dependence is scaled out also through the second moment 
has been established \cite{BCMyR01}.

In spite of its extended use and clear success in many cases, the validity 
of the Boltzmann equation  to describe granular flows has been questioned 
since the early developments of the kinetic theory
of granular systems. One of the first objections raised against its use was 
based on  the tendency of these  systems to form density clusters. Since 
it is only valid in the low density limit, the  Boltzmann equation is 
not suitable to describe the cluster  evolution. This is not a fundamental 
objection, and in fact it can also be raised against the applicability of 
the Boltzmann equation to molecular systems in inhomogeneous states. 
The value of the density limits indeed the applicability of the Boltzmann 
description, but both in the elastic and the inelastic case. 
On the other hand, if we start from an homogeneous, low density, 
initial configuration of a granular gas, under  conditions   that clusters 
will develop eventually, it can be expected that the Boltzmann equation 
will be valid to describe the first stages of the cluster formation, as long 
as  regions with a too large density do not show up. A different objection 
concerns the validity of the molecular 
chaos  hypothesis, which is on the basis of the Boltzmann description. As 
collisions in granular systems tend to make the particle velocities more
parallel, velocity correlations may develop from the early 
stages of the evolution of a granular gas, making the molecular chaos
hypothesis invalid. This problem was directly  addressed by Soto 
et al \cite{SyM01,SPyM01}, who studied the short-range velocity 
correlations in the HCS of a granular fluid, 
concluding that they were not relevant in the low density case. 
Pagonabarraga et al \cite{PTNyE01} also studied the validity of 
the molecular chaos hypothesis in a homogeneous granular system, but
in this case driven by a random force. Again, it was found that, for dilute
systems, deviations from molecular chaos were not significant.
To put this work in a proper context, it should be taken into account that 
the driving force introduced by the authors induced some velocity 
correlations, as  pointed out by them. 

The previous mentioned studies are restricted  to homogeneous states
of a granular gas. They correspond to very special situations, so
the conclusions cannot be extrapolated to more general cases. 
A different situation was investigated recently by Nakanishi
\cite{N03}, who studied the time evolution of the velocity distribution
of a freely evolving granular fluid in conditions such that the HCS 
is unstable.  He found that the kurtosis of the distribution did not remain 
stationary, but evolved  towards the Gaussian
value from the early stages of the evolution of the system, even before the 
clustering instability shows up. He argues that this is in contradiction with
the predictions of kinetic theories based on the Boltzmann equation,
being a clear indication  of the growth of velocity correlations, 
which invalidate the molecular chaos hypothesis. He also found that the 
return to the Gaussian behavior became slower the more dilute the system,
but as changing the density implies changing the ``clustering time'',
it was not clear to the author how the Boltzmann behavior could be recovered.

The object of this work is to investigate the possible differences
in the behavior of the velocity distribution function of a initially
dilute, homogeneous,  granular gas and the predictions of the Boltzmann 
equation  when the system is under conditions such that the HCS is unstable. 
We will compare the results from Molecular Dynamics (MD) simulations at 
different average densities with those from the Direct  Simulation Monte Carlo
(DSMC) method, which is a method to solve numerically the Boltzmann
equation \cite{B94}.

The plan of the paper is as follows: in Section \ref{s2} the kinetic
theory predictions for the HCS are briefly discussed, while the details
of the simulations and the properties to be studied are given
in Sec. \ref{sec3}. The results for the DSMC and MD simulations
are presented in Secs. \ref{sec4} and \ref{sec5}. Some final comments
and discussion are made in Sec. \ref{sec6}.

\section{Kinetic theory for the homogeneous cooling state}
\label{s2}

Let us consider a  a system of $N$ smooth inelastic hard particles,
spheres ($d=3$) or disks ($d=2$), of mass $m$ and  diameter $\sigma$. 
The loss of energy in collisions is  characterized by a 
constant coefficient of normal  restitution $\alpha$, which implies that the  
velocities of two colliding particles $i$, $j$ before and after the collision 
are related by
\begin{eqnarray}
{\bf v}_{i}^{\prime}&=&{\bf v}_{i}-\frac{1+\alpha}{2} ({\bf g}\cdot
\widehat{\mbox{\boldmath$\sigma$}}) \widehat{\mbox{\boldmath$\sigma$}}
\, , \nonumber \\
{\bf v}_{j}^{\prime}&=&{\bf v}_{j}+\frac{1+\alpha}{2} ({\bf g}\cdot
\widehat{\mbox{\boldmath$\sigma$}})\widehat{\mbox{\boldmath$\sigma$}} \, ,
\end{eqnarray}
where the primes denote velocities after the collision, ${\bf g}=
{\bf v}_{i}-{\bf v}_{j}$ is the relative velocity, and $\widehat
{\mbox{\boldmath$\sigma$}}$
a unit vector joining the centers of particles $i$ and $j$ at contact.
Let us notice that $\alpha \leq 1$, and that the value $\alpha=1$ 
corresponds to elastic collisions. The above  rule implies that, 
in each collision, the component of the relative velocity in the direction 
of $\widehat{\mbox{\boldmath$\sigma$}}$ is reduced in a factor $\alpha$:
\begin{equation}
{\bf g}^{\prime}\cdot\widehat{\mbox{\boldmath$\sigma$}}=-\alpha ({\bf g}\cdot
\widehat{\mbox{\boldmath$\sigma$}})
\, .
\label{eq:2}
\end{equation}

The Boltzmann equation describing the evolution of the velocity distribution 
$f({\bf r},{\bf v},t)$ of a system of freely evolving hard particles has 
the form:
\begin{equation}
\left( \frac{\partial}{\partial t}+{\bf v}\cdot{\boldmath \nabla}\right)
f({\bf r},{\bf v},t)=
{\cal J}_{B}[{\bf r},{\bf v}|f({\bf r},{\bf v},t) ] \, ,
\end{equation}
where ${\cal J}_{B}$ is the (inelastic) Boltzmann collision operator:
\begin{equation}
{\cal J}_{B}[{\bf r},{\bf v}|f({\bf r},{\bf v},t) ]=
\sigma^{d-1}
\int d{\bf v}_{1} \int d\widehat{\mbox{\boldmath$\sigma$}}\,
\theta ({\bf g}\cdot \widehat{\mbox{\boldmath$\sigma$}})
({\bf g}\cdot \widehat{\mbox{\boldmath$\sigma$}})
(\alpha^{-2} b^{-1}-1)f({\bf r},{\bf v},t) f({\bf r},{\bf v}_{1},t)\, .
\end{equation}
Here $\theta$ is the Heaviside step function and $b^{-1}$ an operator
transforming velocities ${\bf v}$ and  ${\bf v}_{1}$ to its right
into their precollisional values. As it is the case with elastic particles,
the derivation of this equation is based on the molecular chaos hypothesis,
i.e., the factorization of the two-particle distribution function in
the precollisional sphere:
$f^{(2)} ({\bf x}_{1},{\bf x}_{2},t)= f ({\bf x}_{1},t) f ({\bf x}_{2},t)$,
where ${\bf x}_{i}\equiv\{ {\bf r}_{i},{\bf v}_{i} \}$. If spatial and/or 
velocity correlations are present between colliding particles, 
this factorization does not hold and the assumption of molecular chaos
fails.

When a system of inelastic particles such as the one described evolves 
freely, its  simplest possible state is the HCS. It is a homogeneous state 
with no fluxes, whose temperature $T(t)$, defined as proportional to the 
average kinetic  energy, evolves according to Haff's law\cite{H83},
\begin{equation}
T(t)=\frac{T(0)}{\left(1+\frac{t}{t_{0}}\right)^{2}} \, 
\end{equation}
where $t_{0}$ is the time characterizing the energy decay.

When the system is in the HCS, the Boltzmann equation admits a solution  
$f({\bf v},t)$ that  obeys the scaling law \cite{GyS95,NyE98,BRyC96}
\begin{equation}
f({\bf v},t)=\frac{n_{H}}{v_{0}(t)^{d}} \phi\left(\frac{{\bf v}}{v_{0}(t)}
\right) \, ,
\end{equation}
where $n_{H}$ is the homogeneous density, and $v_{0}$  the thermal velocity
of the system, defined as $v_{0}=\sqrt{2 k_{B} T/m}$, with $k_{B}$
the Boltzmann constant. Therefore, in the HCS all the time dependence of 
the velocity distribution  can be scaled out through its second moment. 
The function $\phi$ is  determined from the 
Boltzmann equation. Although its exact expression is not known, it was
found that it does not deviate much from a Gaussian \cite{BRyC96}. Then,
it is sensible to expand it using the Sonine polynomials, $S^{(j)}$, 
whose explicit expression can be found in   \cite{RyL77}, 
\begin{equation}
\phi({\bf c})=\frac{e^{-c^{2}}}{\pi^{-d/2}} \sum_{j\geq 0} a_{j} 
S^{(j)} (c^{2}) \, , \mbox{\hspace{2cm}}  {\bf c}=\frac{{\bf v}}{v_{0}} \, .
\end{equation}

Normalization and scaling imply $a_{0}=1$ and $a_{1}=0$.  The coefficient 
$a_{2}$ is related to the kurtosis of the distribution
\begin{equation}
a_{2}=\frac{4}{d (d+2)} \left[ \frac{\langle v^{4}\rangle}{\langle v^{2}
\rangle^{2}} -\frac{ d (d+2)}{4} \right] \, , 
\label{eq:a2}
\end{equation}
and its value has been estimated from the Boltzmann equation up to linear 
order in $a_{2}$  \cite{GyS95,NyE98}. Both theoretical calculations and 
numerical simulations  of the Boltzmann equation \cite{BRyC96} show that, 
for not  too inelastic  systems, $a_{2}$ is very small. Deviations from 
gaussianity are important  if we consider the tails of the distribution, 
which are exponential. Again, this has been verified from theoretical 
arguments based on the Boltzmann equation \cite{EyP97,NyE98} and from 
DSMC simulations \cite{BCyR99}.

The time $t_{0}$ characterizing the kinetic energy decay in the HCS has
also been computed by using the Boltzmann equation \cite{BDKyS98},
and its expression reads:
\begin{equation}
t_{0}^{-1}=\zeta^{*}(\alpha)
\frac{4 \pi^{(d-1)/2}}{(2+d) \Gamma(d/2)} \left( \frac{k_{B}
T(0)}{m}\right)^{1/2} n_{H} \sigma^{(d-1)} \, ,
\end{equation}
where $\zeta^{*}(\alpha)$ is a function that depends only on the
coefficient of restitution $\alpha$, and that, for not too
inelastic systems reads \cite{BDKyS98,ByC00}
\begin{equation}
\zeta^{*}(\alpha)\simeq \frac{2+d}{4d}(1-\alpha^{2}) \, .
\end{equation}
In terms of the average number of collisions per particle, $\tau$, Haff's law 
takes the form:
\begin{equation}
T(\tau)=T(0) e^{-2 \zeta^{*} \tau} \, ,
\label{eq:Haff}
\end{equation}
i.e., the kinetic energy decays exponentially with the number of collisions.

It is a well known result \cite{GyZ93,MyY96} that the homogeneous cooling
state of a granular gas is unstable against long wavelength perturbations.
In practice, this implies that, if the system size is larger than
a critical value, $L_{c}$, it will spontaneously develop spatial 
inhomogeneities. The value of the critical size has also been determined 
from the Boltzmann  equation \cite{ByC00,BRyM98}, and it reads:
\begin{equation}
L_{c}=\frac{C_{d} (2+d) \Gamma(d/2)}{2 \pi^{(d-3)/2}} \left(\frac{\eta^{*}}
{2 \zeta^{*}}\right)^{1/2} \lambda_{H} \, .
\label{eq:lc}
\end{equation}
Here, $\lambda_{H}=1/(C_{d} n_{H} \sigma^{(d-1)})$ is the homogeneous
mean free path, $C_{d}=2\sqrt{2}$ if $d=2$ and $C_{d}=\pi\sqrt{2}$ if
$d=3$, and $\eta^{*}$ is a function of  $\alpha$ related to the viscosity
of a granular gas, whose expression can be found in \cite{ByC00}.
Both MD \cite{GyZ93,MyY96} and DSMC 
\cite{BRyC99} simulations of freely evolving granular gases show that,
in the development of inhomogeneities, first velocity vortices appear in 
the system and then the density becomes also very inhomogeneous. Of course,
once the instability has set in, the velocity distribution of the system
will no longer be the one of the HCS. Therefore, one expects  the
scaling to fail and $a_{2}$ to become time dependent. The question is whether
the Boltzmann equation is valid to describe how the system departs
from the HCS. It might happen that, as pointed out by Nakanishi \cite{N03},
the development of velocity correlations are very important from the
early stages of this departure, and then the Boltzmann equation fails. 
This is the main point we want to clarify in this work.

\section{Computer Simulations}
\label{sec3}

Given that we want to check the validity of the Boltzmann equation
to describe granular gases, we will compare MD simulations
of  initially homogeneous, low density, granular systems with the predictions
of the Boltzmann equation. Since the analytical solution of the
latter is not known, we will use the DSMC method developed by Bird \cite{B94} 
to construct numerical solutions of the Boltzmann equation under 
the desired conditions. It must be remembered
that this method mimics the dynamics behind the Boltzmann equation  
by uncouplig the free flow of particles and collisions during a small
enough time interval. Besides, collisions are always treated as if there
were no correlations between colliding particles. In practice, this
implies that space is divided in cells of size smaller than the
mean free path, and particles within a cell collide with a probability
proportional to their relative velocity. Technical details about the DSMC
method have been extensively discussed in the literature \cite{B94,G00}.
The only point we want to stress is that, when using this method, the
simulated system is {\em by definition} in the low density limit,
where the Boltzmann equation is supposed to apply, no
matter how many simulated particles there are in a given space region. 

	The simulations we will present in the following  correspond to a
two-dimensional system of freely evolving hard disks in a square box
of side $L$, larger than the critical size $L_{c}$ given by Eq. (\ref{eq:lc}).
Periodic boundary conditions  will be considered in all the cases. 
In the MD simulations,  the event driven algorithm \cite{AyT87} will
 be used. For our purposes, it is important to study the effect of lowering
the density on the evolution of the system. For that reason, two
different, low values, of the density will be studied in the MD
 simulations, namely $n_{H}=0.05\sigma^{-2}$ and
 $n_{H}=0.0125\sigma^{-2}$. Nevertheless, it must be noticed that,
for a given $\alpha$, changing the density implies changing
the critical size of the system, as it follows from (\ref{eq:lc}).
As a consequence, if the systems have the same size $L$,  
the characteristic time governing 
the growth of instabilities will be also changed, and the departure from 
the HCS will be faster in the denser system. Therefore, if we want to have 
the same ``distance to stability'' in systems with different densities,
their  sizes must be changed correspondingly, so $L/\lambda_{H}$
remains the same. The same value of this scaled
size should be considered in the DSMC simulations, again to expect the
same characteristic time governing the departure from the HCS.

	Moreover,  we have used a fixed value  of the  restitution 
coefficient, $\alpha=0.9$, which is not too inelastic,
but lies outside of what  can be considered the quasi-elastic region. 
For this value of $\alpha$, Eq. (\ref{eq:lc}) leads to 
$L_{c}\simeq 23.36\lambda_{H}$. As we want the 
system to be well inside the unstable region, we have chosen the system
size to be $L=80\lambda_{H}$. Then, for $n_{H}=0.05\sigma^{-2}$ the
number of particles in the system was $N=16000$, while for 
$n_{H}=0.0125\sigma^{-2}$, $N=64000$. In the DSMC simulations, 
$N=2.048\cdot 10^{6}$ particles were considered, but it must be reminded 
that this number has only a statistical meaning. In all the cases, we started 
with the particles homogeneously distributed in the system
and with a Gaussian velocity distribution. This situation was
let to evolve with elastic collisions during several collisions
per particle, until the system was equilibrated. Then, the inelasticity
was switched on, and this time is taken as the origin, $t=0$, of
our simulations. The initial elastic period ensures that the structure of 
the fluid at $t=0$ is the equilibrium one. 

\subsection{Studied properties}

	As the aim of the paper is to investigate the departure of the
velocity distribution of the system from the HCS distribution, this will
be one of the properties to be studied in the simulations. To be more
precise, we will compute the second $\langle v^{2} \rangle$ and fourth 
$\langle v^{4} \rangle$ velocity moments of the total velocity 
distribution, and from them we will compute also $a_{2}$, given by
Eq. (\ref{eq:a2}). As far as the system stays in the HCS, $a_{2}$
will be constant. It is important to notice that, as
in Ref. \cite{N03},  we will study the
global velocity distribution of the system, even if inhomogeneities 
have appeared,  causing  local averages of the physical quantities
to be quite different from their global values.

      The growth of inhomogeneities will be controlled by following
the evolution of the density $n({\bf r},t)$, and velocity 
${\bf u}({\bf r},t)$, fields. To compute them, the system is divided
into 30x30 square cells of size $l_{c}\sim 2.7\lambda_{H}$, and properties
are averaged in each cell. Again, while the system remains in the HCS,  
$n({\bf r},t)$ should be constant and ${\bf u}({\bf r},t)=0$ (apart
from  statistical noise). Once the instability develops, vortices
and also density inhomogeneities will show up. The question remains of how
the departure of the velocity distribution from the HCS one is related to
the development of these inhomogeneities.

	The possible failure of the molecular chaos hypothesis 
will be investigated by the study of several collisional averages,
i.e., averages over colliding pairs of particles. 
These  averages contain information
about the two-particle distribution of colliding particles, which is
the function whose factorization is assumed in the molecular chaos hypothesis.
The first of the collisional averages we will consider is the pair 
distribution function at contact, $g(\sigma^{+})$. In a elastic system,
if there are no correlations between colliding particles, 
as it is in fact assumed by the Boltzmann equation, $g(\sigma^{+})=1$. 
In the case of a system 
of inelastic disks, the pair distribution function at contact  can be easily
computed by using \cite{L01} (notice that there is a miss-print in the expression
for $g(\sigma^{+})$ provided in the cited reference):
\begin{equation}
g(\sigma^{+},t)=\frac{1+\alpha}{\alpha}\frac{1}{N n \pi\sigma}
\frac{1}{\Delta t} \sum_{\gamma \in \Delta t} \frac{1}{ 
|\widehat{\mbox{\boldmath$\sigma$}} \cdot {\bf g}_{\gamma}|}
\theta(- \widehat{\mbox{\boldmath$\sigma$}} \cdot {\bf g}_{\gamma}) \, ,
\label{eq:g}
\end{equation}
where we are summing over all collisions $\gamma$ taking place in the interval
$\Delta t$, and the $\theta$ function implies that we are using the
pre-collisional values of the quantities in the sum. When the system is in 
the HCS, it has been found \cite{L01,SyM01} that $g_{HCS}(\sigma^{+})$ 
does not depend on time, and takes the value:
\begin{equation} 
g_{HCS}(\sigma^{+})=\frac{1+\alpha}{2\alpha} g_{0}(\sigma^{+})\, ,
\label{eq:ghcs}
\end{equation}
with $g_{0}$ the equilibrium elastic pair distribution function at contact, 
that in  the two dimensional case is accurately given by \cite{H75}
\begin{equation}
g_{0}(\sigma^{+})=1+\frac{\pi (25-4 n \pi) n}{4 (4-n\pi)^{2}} \, .
\end{equation}

In our simulations, time will be discretized  so the  average over collisions 
in (\ref{eq:g}) will be done not over the whole simulation
time, but over collisions occurring in a given interval. In this 
way, we can study the evolution of the collisional averages and,
in particular, of the pair distribution function at contact. Needless to
say, $g(\sigma^{+},t)$ will only be computed in the MD
simulations, as in the DSMC method its value is given.

Another test of the validity of the molecular chaos hypothesis 
will be provided by the probability distribution of the impact parameter,
$b=\frac{\sigma}{2} \sin \chi$, where $\chi$ is the angle formed by the 
impact relative velocity ${\bf g}$ and the unit vector 
$\widehat{\mbox{\boldmath$\sigma$}}$. 
In a system of hard disks, and if there are no correlations, the
impact parameter is uniformly distributed. Soto et al \cite{SyM01}
measured this distribution in the HCS and found no deviations from
uniformity. Here, we will consider the same discretization discussed
above to construct the distribution of impact parameters in each time
interval, and study its possible time dependence. Again, the impact
parameter distribution will be studied only in the MD
simulations, as the DSMC method assumes its uniformity.

	The possible velocity correlations will also be studied. A quantity
that has been used to study them  is
the collisional average $\Gamma(t)$ defined as
\begin{equation}
\Gamma(t)=\frac{1}{{\cal N}_{\gamma}} \sum_{\gamma\in\Delta t} 
\frac{{\bf c}_{i}^{\gamma}\cdot {\bf c}_{j}^{\gamma}}
{|{\bf c}_{i}^{\gamma}-{\bf c}_{j}^{\gamma}|} 
\theta(- \widehat{\mbox{\boldmath$\sigma$}} \cdot {\bf g}_{\gamma}) \, ,
\label{eq:gam}
\end{equation}
with ${\cal N}_{\gamma}$ the number of collisions  in the interval $\Delta t$,
and $i$, $j$ the particles involved in collision $\gamma$. If there is
neither  macroscopic velocity field nor velocity correlations in the system, 
it is $\Gamma=0$. Therefore, in the HCS, non-zero values of $\Gamma$ are a 
clear signal of the presence of velocity correlations in the system. When
the system leaves the HCS, there is another possible reason for these
non-zero values. The build-up of velocity vortices implies 
that $u({\bf r},t)$ is
different from zero, and then, even if there are no velocity correlations,
$\Gamma$ will be different from zero. Let us notice that $\Gamma(t)$
can be measured both in Molecular Dynamics and DSMC simulations.

	Finally, the distribution of the angle $\phi$ formed by the 
velocities ${\bf v}_{i}$,  ${\bf v}_{j}$, of colliding particles, will also 
be computed. We are not aware of any  analytic expression for this quantity 
in the elastic case, and, therefore,  we will use the initial, 
elastic part, of the 
simulations to obtain the elastic distribution of the $\phi$ angle. 
The parallelization mechanism inherent to inelastic collisions may
cause deviations from this behavior, even in the HCS. In any case, 
this distribution shows if there is a predominance of collisions
between particles moving more or less parallel in the system. Again, this
distribution will be measured both in MD and DSMC simulations. 

	It is important to remark the different physical information
behind the distributions $P(b,t)$ and $P(\phi,t)$. The former 
is related to the spatial distribution of the incident flux over
colliding particles, while the latter contains only information
about the relative direction of the velocities of colliding particles.

\section{Results}
\label{sec4}  

The results we will present correspond, as we have already stated,
to MD and DSMC simulations of a freely evolving system of hard disks
with $\alpha=0.9$ and $L=80\lambda_{H}$, i. e., in conditions such
that the HCS is well inside the unstable region. In the following,
the mass $m$ of the particles will
be used as the unit of mass, and the initial kinetic energy as
the unit of energy. The results we will present have been averaged
over several trajectories in all the cases in order to improve
the statistics. Besides, the time evolution of 
the system will be expressed in terms of the number of collisions per 
particle, $\tau$. As the system is prepared in an initially homogeneous 
situation, we expect it  to stay  in the HCS for a transient period, 
until the instability sets in. From that moment,
the different physical properties will depart from their HCS values.

In Fig.\ \ref{fig:1} we have plotted the evolution of the average
kinetic energy per particle, $\langle v^{2} \rangle/2$,  in the system. 
This quantity is proportional to the granular temperature as far as 
there is no macroscopic velocity field, i.e., while ${\bf u}=0$. Also
plotted is Haff's law, Eq. (\ref{eq:Haff}), describing the evolution
of this property in the HCS. In all the cases there is an initial
period in which the energy  follows Haff's law and,
after that, the average kinetic energy of the system decays
slower than in the HCS. The departure form Haff's law occurs
sooner in the MD simulations than in the DSMC one, but it is important
to notice that, if we compare the two MD simulations, it  occurs sooner 
in the case of larger density. Therefore, it seems that, although
there are differences between the behavior of a finite density
granular gas and the predictions of the Boltzmann equation,
they are smaller the smaller the density. Then, it seems sensible
to expect that the Boltzmann equation predictions provide indeed
the correct picture in the analytic low density limit, $n\rightarrow 0$.

The evolution of $a_{2}$ is plotted in  Figs.\ \ref{fig:2}. In
Fig. \ \ref{fig:2}(a), the initial evolution of this quantity is shown: 
there is an initial, very fast, decay from the initial Gaussian, 
$a_{2}=0$, value to the HCS one, followed by a steady period while the velocity
distribution of the system remains in the HCS. The complete time
evolution of the system is given in Fig. \ \ref{fig:2}(b). Again, in all
the cases we observe a similar behavior. After the initial steady period, 
$a_{2}$ grows, reaches a maximum, and afterwards it decays in time. 
The approximate
duration of the steady period, $\tau_{st}$, is $\tau_{st}\sim 15$ for
$n_{H}=0.05\sigma^{-2}$,  $\tau_{st}\sim 30$ for $n_{H}=0.0125\sigma^{-2}$
and $\tau_{st}\sim 50$ in the DSMC simulation. The position of the
maximum, $\tau^{*}$, is $\tau^{*}\sim40$ for $n_{H}=0.05\sigma^{-2}$,
$\tau^{*}\sim 50$ for $n_{H}=0.0125\sigma^{-2}$, and $\tau^{*}\sim 70$
in the DSMC simulation. The height of the maximum is also larger the
larger the density. A first conclusion that can be extracted from 
Figs.\ \ref{fig:2} is that the qualitative behavior of $a_{2}$ 
that follows from the Boltzmann equation is the same found in the
MD simulations. Besides, the differences between the MD simulations and
the DSMC one are smaller the smaller the density in the formers,
showing again a tendency to the Boltzmann behavior in the limit
of very low density. 

Comparison of Figs.\ \ref{fig:1} and \ref{fig:2} is interesting in
order to determine the sensitivity of Haff's law to the exact shape
of the velocity distribution function. It is found that, for $\tau=\tau_{st}$,
when $a_{2}$ begins to depart from the steady value, the temperature
takes the value predicted by Haff's law in all the cases, as it is
expected. What it might be surprising is that, at the maximum $\tau^{*}$, the 
deviations from the Haff value are relatively small, the temperature being
at most $1.3$ times the Haff's value. This is in agreement with simulations
of homogeneous systems of  inelastic hard disks, that show that the evolution 
of the temperature follows quite approximately Haff's law, unless the
fourth moment of the velocity distribution is very large as compared
with the second one \cite{byr03u}.

Once the validity of the Boltzmann description as a limit to the behavior
of a granular gas for vanishing density has been established (at least
with regards to the behavior of the second and fourth moment
of the velocity distribution), a natural emerging question is which is 
the mechanism  taking the velocity distribution out of the HCS shape.
With that purpose, the evolution of the density and velocity
fields have been studied. It must be noticed that, as we expect them to be
inhomogeneous, and their spatial distribution changes from one realization to
another,  these fields cannot be averaged over different runs.

Let us consider first the evolution of the density field. In all the cases 
we observe that the system remains homogeneous during what we have called the
steady period, $\tau_{st}$. Between $\tau_{st}$ and $\tau^{*}$ 
small density inhomogeneities begin to show up, but they are
not very significant. The situation changes from $\tau^{*}$ on: there
is a very fast growth of density inhomogeneities, and in all the
cases, elongated clusters of particles are formed, similar to those
observed in previous studies of freely evolving granular systems.
This behavior of the density field is quantified in Figs.\ \ref{fig:3}
and \ref{fig:4}. In the first of the them, the evolution of the dispersion
of the density fluctuations, $\delta=\sqrt{\langle (n({\bf r},t)-n_{H})^{2}
\rangle}$ is plotted. Let us notice that this quantity is non-zero even
in a homogeneous state, due to statistical noise, and its value in
the homogeneous state depends on the number of particles per cell,
which is different in each of the simulations. For that reason 
$\delta$ has been scaled with its value in the elastic part of the simulation,
$\delta_{el}$. It is observed in the figures that, 
in all the cases, $\delta$ has not increased much
over its elastic value at $\tau=\tau^{*}$, but from then on there is a very
sharp growth of the density fluctuations as a consequence of the formation
of clusters in the system. A similar behavior is exhibited by the maximum
value of the density, $n_{M}$, which is plotted in 
Fig.\ \ref{fig:4}, scaled with the homogeneous  density. 
While the system is in a homogeneous state, $n_{M}$ is a 
measure of the statistical noise. Let us remember that the systems
were divided in 30x30 cells to compute the hydrodynamics fields: as
the number of particles is smaller in the denser system, the noise
in the fields will be larger, and that is the reason why, in the
homogeneous part of the evolution, $n_{M}/n_{H}$ is larger the
larger the density. It must be also pointed out that, once
the clustering begins, there is a time interval in which the growth of
$n_{M}$ can be fitted to an exponential, 
\begin{equation}
n_{M} \sim n_{H}+C e^{s_{n} (\tau-\tau_{1})} \,
\end{equation}
where $C$ and $\tau_{1}$ are constants whose value is not relevant for the 
present discussion, and $s_{n} \sim 0.13$ for
$n=0.05\sigma^{-2}$,    $s_{n} \sim 0.163$ for $n=0.0125\sigma^{-2}$, and
$s_{n} \sim 0.167$ in the DSMC simulation. Again, the discrepancies
between the Boltzmann behavior and the one at finite density are smaller
as we lower the density, being  the growth rate almost the same
for the lower density case and the DSMC simulation.

	In a freely evolving granular system, it has been shown that,
when using $\tau$ as the time variable, 
the growth rate of the scaled transversal velocity mode is 
\begin{equation}
s_{\perp}=\zeta^{*}-\frac{\eta^{*}}{2} k^{2} \, ,
\end{equation}
where $k$ is a non-dimensional wave-vector defined as $k=2\pi/(\sqrt{\pi}
n_{H}\sigma  l)$, with $l$ the wave-length of the perturbation.
In a simulation, the maximum allowed wavelength, which
is the one that grows faster,  is $l=L$. For the values of the parameters
used in this work,  the theoretical prediction is  
$s_{\perp}\sim 8.64\cdot 10^{-2}$. It has been
argued \cite{GyZ93,BRyC99b} that the density inhomogeneities growth in a 
freely evolving granular system is a consequence of the non-linear coupling 
between the transversal velocity mode and the other hydrodynamic
fields, in particular, the density.
If that is indeed the case, the growth rate of the density, at least
in its first stages, should be $2 s_{\perp}$. Here, $2 s_{\perp}
\sim 0.173$,  which is very close to  the value of $s_{n}$ found in the
DSMC and in the MD lowest density simulation, confirming once again
the picture described above. This agreement indicates the hydrodynamical 
character of the density fluctuations in the clustering regime.

The growth of the scaled velocity field has also been followed in our 
simulations, and we found that vortices begin to develop quite soon
in the system. To quantify this, we introduce $\delta_{u}$, the scaled 
average value of ${\bf u}^{2}$, as $\delta_{u}=\langle n({\bf r},t)
 {\bf u}^{2}({\bf r},t) \rangle$.
Again, due to statistical noise, $\delta_{u}$ will be different
from zero when computed from a simulation, even if there are no
fluxes. For that reason, in Fig.\ \ref{fig:5} we have plotted
$\delta_{u}$ scaled with its value in the elastic part of the
simulation, $\delta_{u}^{el}$. All the simulations show that the velocity 
field grows from the very early stages of the evolution. In fact, there is
a first part of the growth of $\delta_{u}/\delta_{u}^{el}$, which
can be quite well approximated by an exponential, that is common to
all the simulations. After that, there is a saturation effect that translates
into deviations from the exponential growth for larger times, occurring
sooner the larger the density. In fact, the saturation occurs
for times of the order of $\tau^{*}$, which is the time when the
clustering is triggered.

\section{Collisional averages}
\label{sec5}

The evolution of the collisional averages in the system has been computed by 
discretizing the time in intervals $\Delta \tau=5$. Properties are averaged
over collisions taking place in each interval. Also, all the results 
have been averaged over several runs.

In Fig.\ \ref{fig:6} we have plotted the evolution of the pair correlation 
function  at contact, $g(\sigma^{+})$, from the MD simulations. Also 
included is the theoretical  prediction in the HCS for the two values 
of the density displayed in the figure, calculated from Eq. (\ref{eq:ghcs}). 
In both cases it is found that,  
when the velocity distribution begins to  depart from the HCS one, i.e., 
at $\tau=\tau_{st}$, the pair correlation function takes still
the HCS value and, in fact,  deviations from it are quite small
even at $\tau=\tau^{*}$. After that, there is a very fast increase
of $g(\sigma^{+})$ due to the formation of clusters in the system.
Therefore, the increase in $a_{2}$ shown in Fig.\ \ref{fig:2} is not
due to the development of positional correlations of colliding
particles. These correlations do appear, but at rather later times. 
We have also investigated the evolution of the impact parameter
distribution, $P(b,t)$ and found that in none of the MD simulations
discussed in this work it deviated significantly from uniformity for 
the times shown in this manuscript. 

	Velocity correlations were investigated through
the behavior of $\Gamma$ defined in Eq. (\ref{eq:gam}), 
and of the distribution function of the angle
formed by the velocities of colliding particles, $\phi$. In Fig.\ \ref{fig:7},
the evolution of $\Gamma$ for the three simulations is shown.
It must be noticed that, in a finite system, even if there are no 
correlations,  $\Gamma$  takes a finite  
value that depends on the number of particles, which is different in the
three cases. So, even in the elastic part of the simulation, $\Gamma$ 
is different in each case, being larger in the denser system, as it has 
less particles. Besides, the value of $\Gamma$ is different in a elastic
system and in a inelastic one in the same conditions. Therefore, 
when the inelasticity is switched on at $t=0$, there is a very fast 
increase of $\Gamma$ to its HCS value. Then, there is a quite steady
period that lasts longer the lower the density, followed by
an almost exponential increase of  $\Gamma$. It is found that the
growth rate in this period depends on the density, becoming larger
as $n$ decreases. Finally, there is a slowing down in the increase of 
$\Gamma$, and it seems that it saturates in the end to a value 
which is larger  the larger the density. Also the slowing down begins
sooner the larger the density. In any case, we find again that by lowering
the density in the MD simulations,  the behavior of the system 
approaches the one predicted by the Boltzmann equation. One could be
tempted to conclude that the increase of $\Gamma$ is due to the  development
of correlations between velocities of  colliding particles, this
happening from the early stages of the evolution of the system.
But one must be cautious when interpreting this result. Let us remember  
that in our system  a velocity field is also being built up from the 
beginning of the evolution, as it was shown in Fig.\ \ref{fig:5}, 
and this  leads
to an increase in the value of $\Gamma$ that has nothing to do with the
failure of the factorization of the two-particle distribution function
of colliding particles, i.e., with a violation of the molecular
chaos hypothesis. In fact, the behavior of $\Gamma$ displayed in 
Fig. \ref{fig:7} is quite reminiscent of the one of the fluctuations of 
the velocity field. Also, in the DSMC simulation, it
must be taken into account that  the molecular chaos hypothesis
is assumed in the very basis of the algorithm, so the increase 
of $\Gamma$ in that case cannot be due to the presence of pre-collisional
velocity correlations. We conclude then that  the growth of $\Gamma$ 
displayed in Fig. \ref{fig:7} is due to the instability of the scaled 
transversal velocity mode, that leads to the formation of vortices and, as
a consequence, the velocities of colliding particles become more and
more parallel.

The existence of the parallelization mechanism is investigated also 
by studying the evolution of the distribution of the angle formed by
the velocities of colliding particles, $P(\phi)$. In Fig. \ref{fig:8}
this distribution is plotted at different times for the MD simulation
with $n=0.0125\sigma^{-2}$. We have included the 
distribution at $\tau=0$, which is constructed from the elastic part of
the simulation. While the system remains in the HCS, the distribution
of the $\phi$ angle is indistinguishable from the elastic one, so there 
are no apparent angle correlations in this part of the evolution of 
the system. The situation changes at $\tau\sim\tau_{st}$  ($\tau_{st}\sim 30$ 
in this case), when deviations from the elastic distribution begin to show up. 
The relative number of collisions with larger relative angles of the 
velocity begin to  decrease with respect to the elastic case. As time
proceeds, this tendency becomes stronger, and for the final times
considered in the simulation, most of the collisions correspond to
particles that are moving almost parallel. The distortion of the
angle distribution begins precisely at the time when the kurtosis
of the total velocity distribution begins to depart from the HCS value.
It seems sensible to conclude then that the behavior of the velocity
distribution of the system is a consequence of the parallelization
mechanism, which is inherent to inelastic collisions, and that,
when the system is unstable, induces a collective behavior of
the velocities of particles, that translates into a departure
of the velocity distribution from the HCS one.
  
 The qualitative behavior of $P(\phi)$ discussed for  $n=0.0125\sigma^{-2}$
also holds for the larger density MD simulation and for the
DSMC one. Of course, the deviation from the elastic distribution
occurs sooner the larger the density, but it is also present in the
Boltzmann description. To have a clear picture of the parallelization
of the velocities of colliding particles in the three cases, in
Fig. \ref{fig:9} we have plotted the evolution of $\langle \phi \rangle$
in the three simulations. The value of this quantity in an equilibrium,
elastic fluid, is $\langle \phi \rangle \simeq  0.59\pi (\sim 107^{o})$.   
In the three cases, the system remains with the elastic
value of $\langle \phi \rangle$ during what we called the steady period, 
and then it begins to decrease with time, indicating
that there is a tendency to have more collisions with velocities that form
a small angle. It must be also remembered that, in the MD simulations, 
the impact parameter distribution did not show deviations from uniformity
for the times considered in this work.

\section{Discussion}
\label{sec6}
 
The simulation of a system on freely evolving hard disks in conditions
such that the HCS is unstable shows that the Boltzmann description
provides a valid picture for the behavior of a low density granular
gas. This has been established by showing that there are no qualitative
differences between the results obtained in low density MD simulations
and those that follow by using the DSMC method. It has been shown that, when
the density is lowered in the suitable way in the MD simulations, i.e., leaving
the characteristic time for the development of instabilities unchanged, 
the behavior of the system tends to the Boltzmann one. 
Nevertheless, it is also true that the deviations from the Boltzmann
behavior are larger  in a system in these conditions of
instability than in a stable one, for the same values of the
density and restitution coefficient. The reason for this seems to be
that, when the HCS in unstable, there is a parallelization
of velocities of colliding particles that is more efficient the higher 
the density, although  it is also present in the Boltzmann description.

	The role of the spatial correlations between colliding
particles has been investigated 
in this work by studying the pair distribution at contact and the 
probability distribution of impact parameters. The former only
deviate from the HCS value when density inhomogeneities
are already developed in the system, while the latter remains
always uniform.  This implies that, in
a low density granular gas, the development of spatial correlations 
does not play a significant role in the early departure from the HCS. 

The above results seem to indicate that, when trying to extend
the inelastic Boltzmann equation to finite higher density, the effect of
velocity correlations between colliding particles must be
incorporated. At least, in the physical situation considered in
this paper, velocity correlations become important quite before
the system develops significant spatial correlations, as measured
by the pair distribution function at contact. It is sensible to
expect that this effect increases as the inelasticity of the
system increases. If this picture were right, kinetic equations for 
inelastic dense gases should not be based on Enskog-like equations,
taking into account only spatial correlations, but new approximations
incorporating the effect of velocity correlations in the precollisional 
sphere are needed. This implies a rather strong departure from the
traditional methods of kinetic theory for elastic systems.

\acknowledgments

We acknowledge partial support from the Ministerio de Ciencia y Tecnolog\'{\i}a
(Spain) through Grant No. BFM2002-00303 (partially financed by
FEDER founds).

\begin{figure}
\caption{Evolution of the average kinetic energy  for the simulations
discussed in the text. The units are chosen such
that the initial value is equal to one. Also shown is the theoretical
prediction for the HCS, i.e., Haff's law. Time is measured in average
number of collisions per particle.}
\label{fig:1}
\end{figure}

\begin{figure}
\caption{Time evolution of the dimensionless coefficient $a_{2}$ 
for the simulations discussed in the main text. The short time behavior,
(a),  and the complete evolution, (b),  are displayed in different plots for 
the sake of clarity.}
\label{fig:2}
\end{figure}

\begin{figure}
\caption{Evolution of the average value of the density fluctuations,
scaled with their value in the elastic case, for the simulations
discussed in the main text.}
\label{fig:3}
\end{figure}

\begin{figure}
\caption{Evolution of the  maximum value of the scaled density for the
simulations discussed in the main text. The symbols are from the 
simulations, and the dotted line is just a guide to the eye.}
\label{fig:4}
\end{figure}

\begin{figure}
\caption{Evolution of the fluctuations of the velocity field, scaled
with its value in the elastic case, for the simulations discussed 
in the main text.}
\label{fig:5}
\end{figure}

\begin{figure}
\caption{Evolution of the pair distribution function at contact, 
$g(\sigma^{+})$, obtained in the two MD simulations discussed
in the main text. The horizontal lines are the theoretical prediction
for this function in the HCS at $n=0.0125\sigma^{-2}$ and 
$n=0.05\sigma^{-2}$, from bottom to top.}
\label{fig:6}
\end{figure}

\begin{figure}
\caption{Evolution of  $\Gamma$ for the simulations discussed in the
main text.}
\label{fig:7}
\end{figure}

\begin{figure}
\caption{Probability distribution of the angles of the 
velocities between colliding
particles, P($\phi$), from the MD simulation with $n=0.0125\sigma^{-2}$ .}
\label{fig:8}
\end{figure}

\begin{figure}
\caption{Evolution of the average value of the angle between the velocities
of colliding particles, $\langle \phi \rangle$, for the simulations discussed 
in the main text.}
\label{fig:9}
\end{figure}

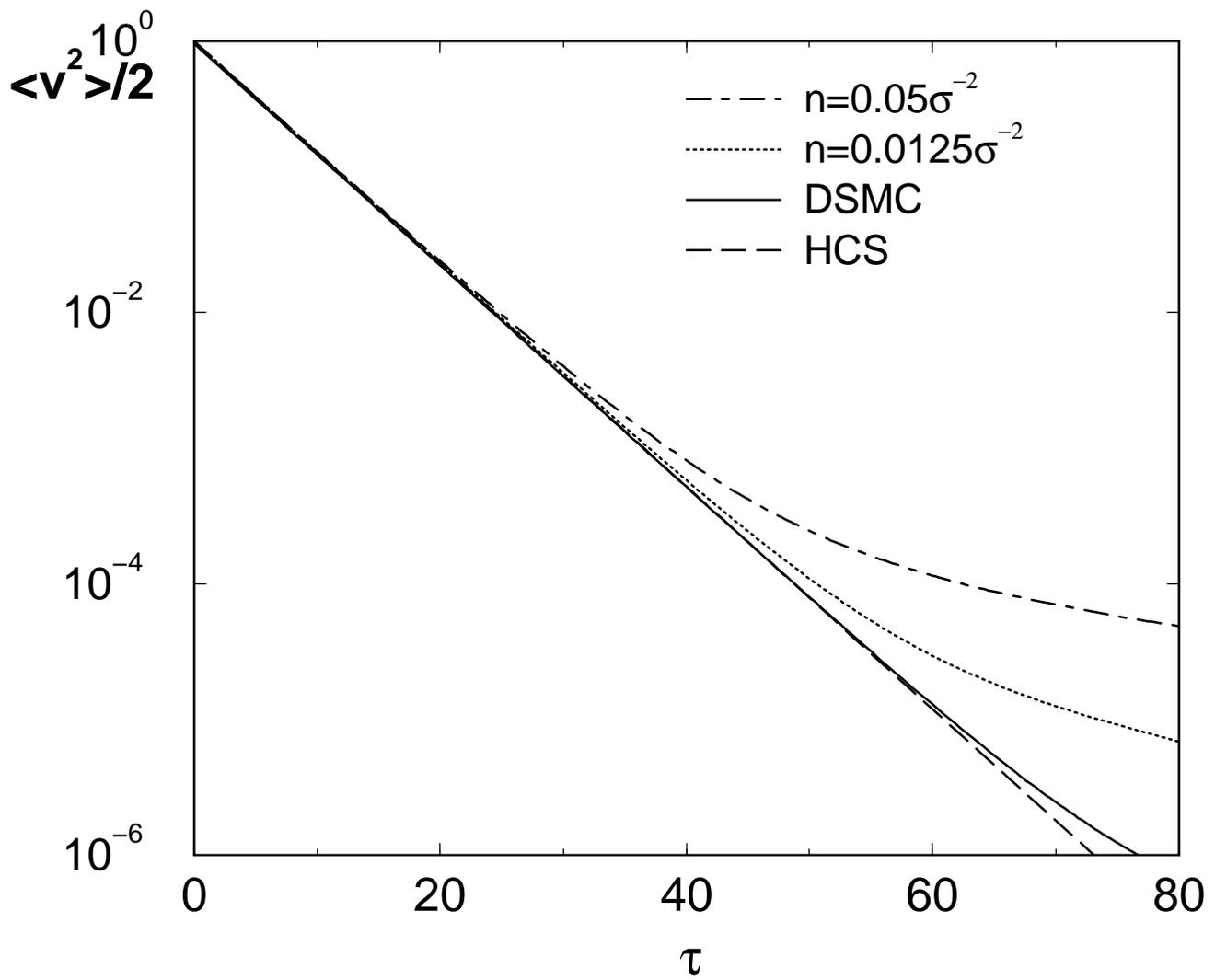
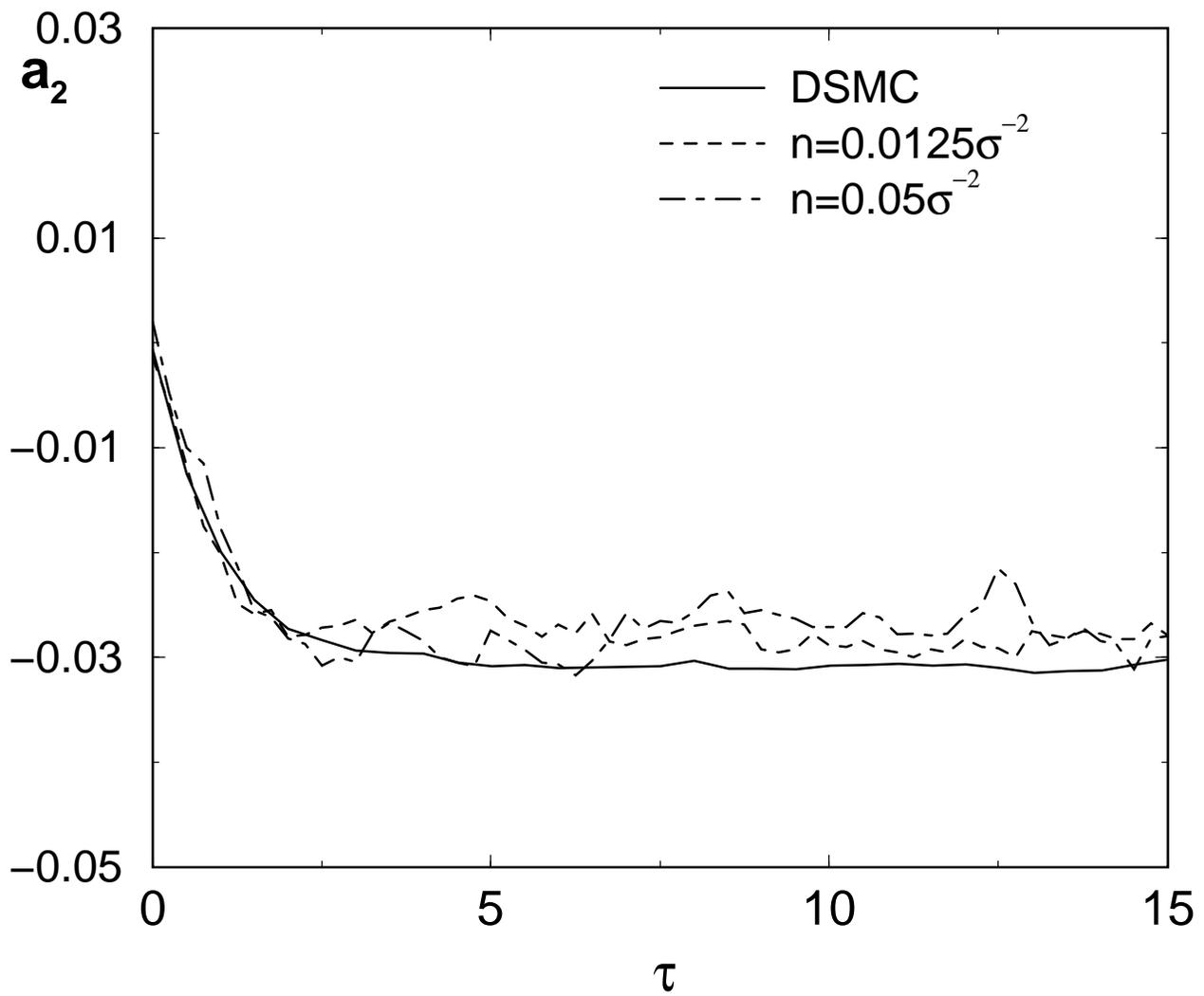
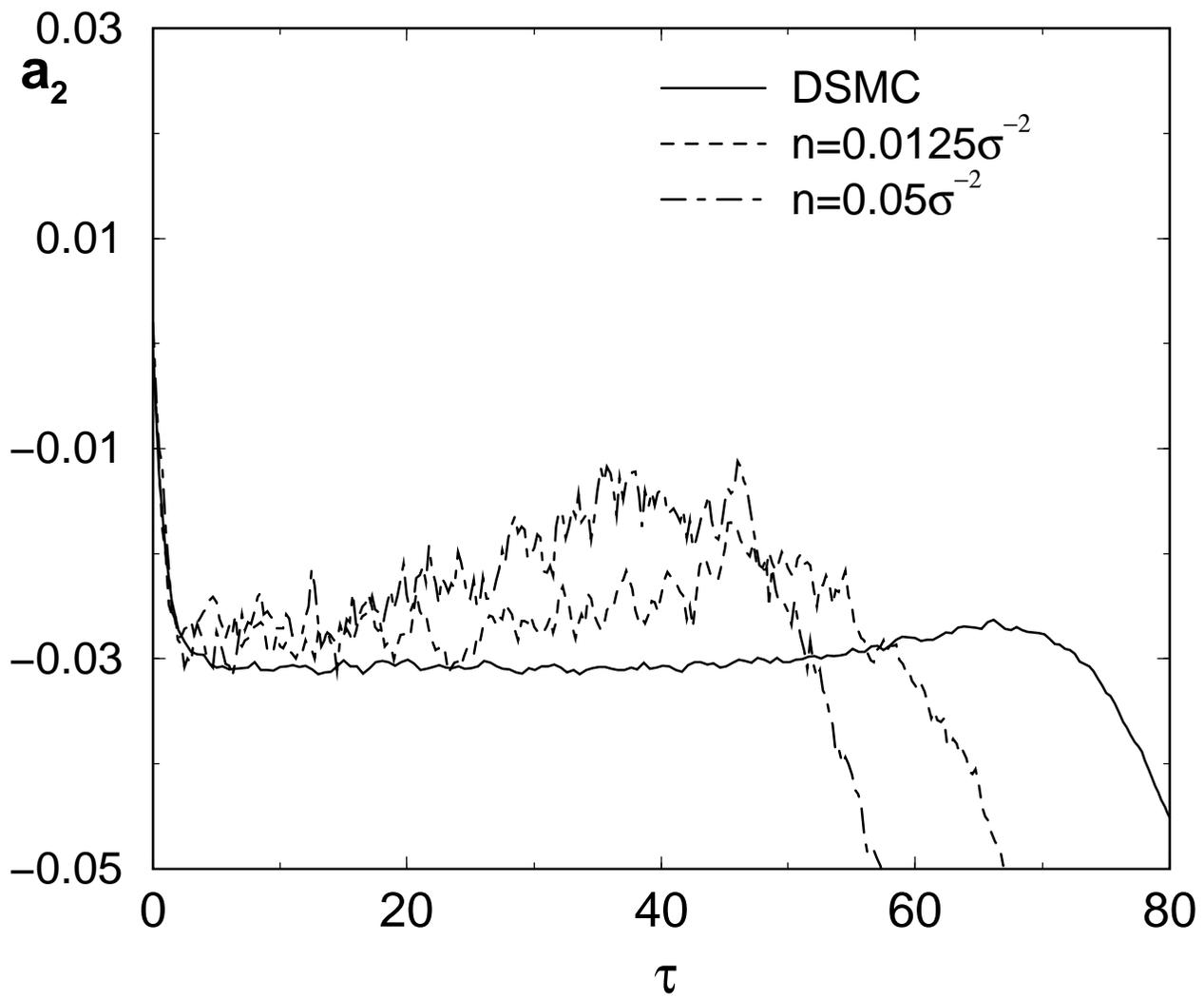
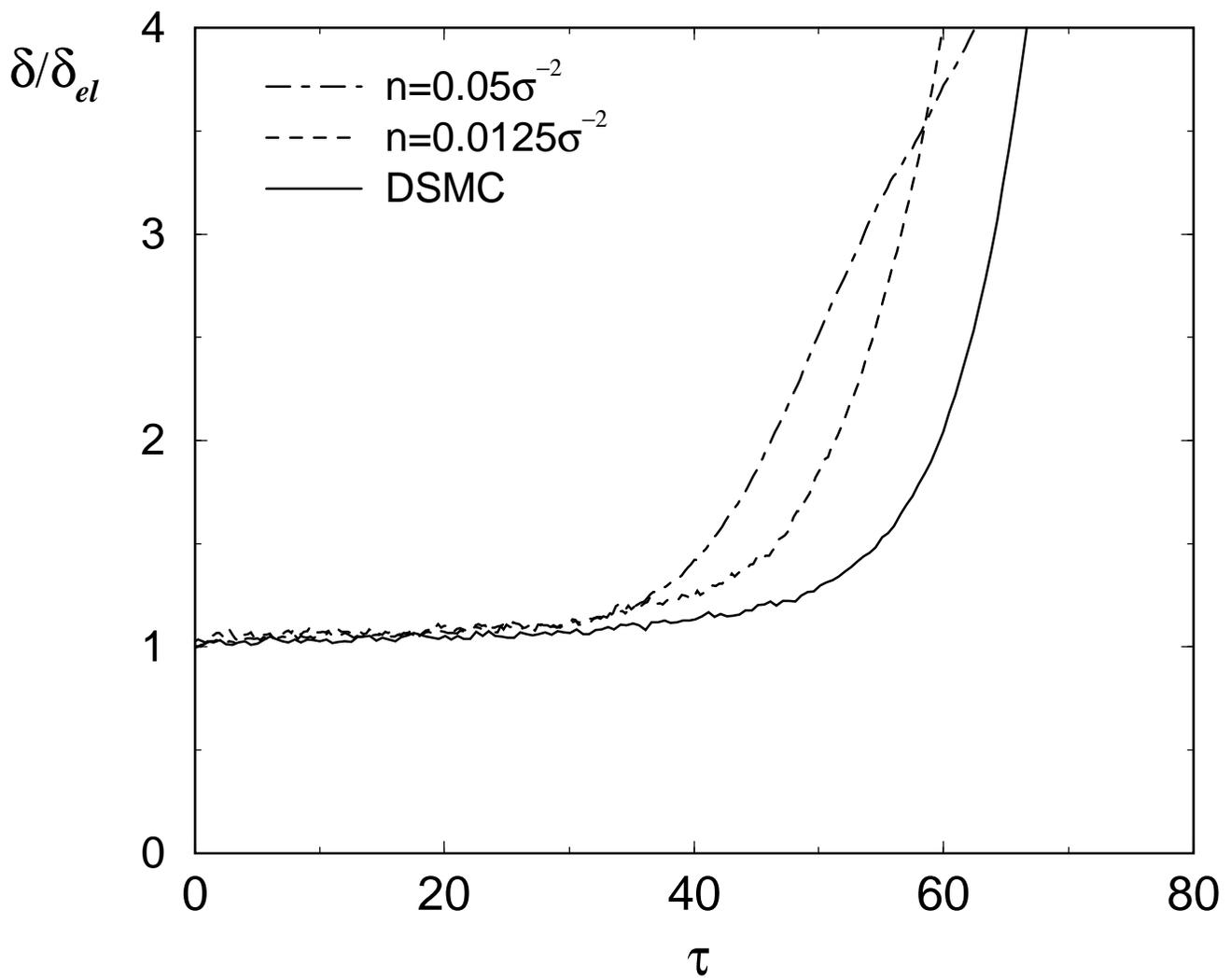
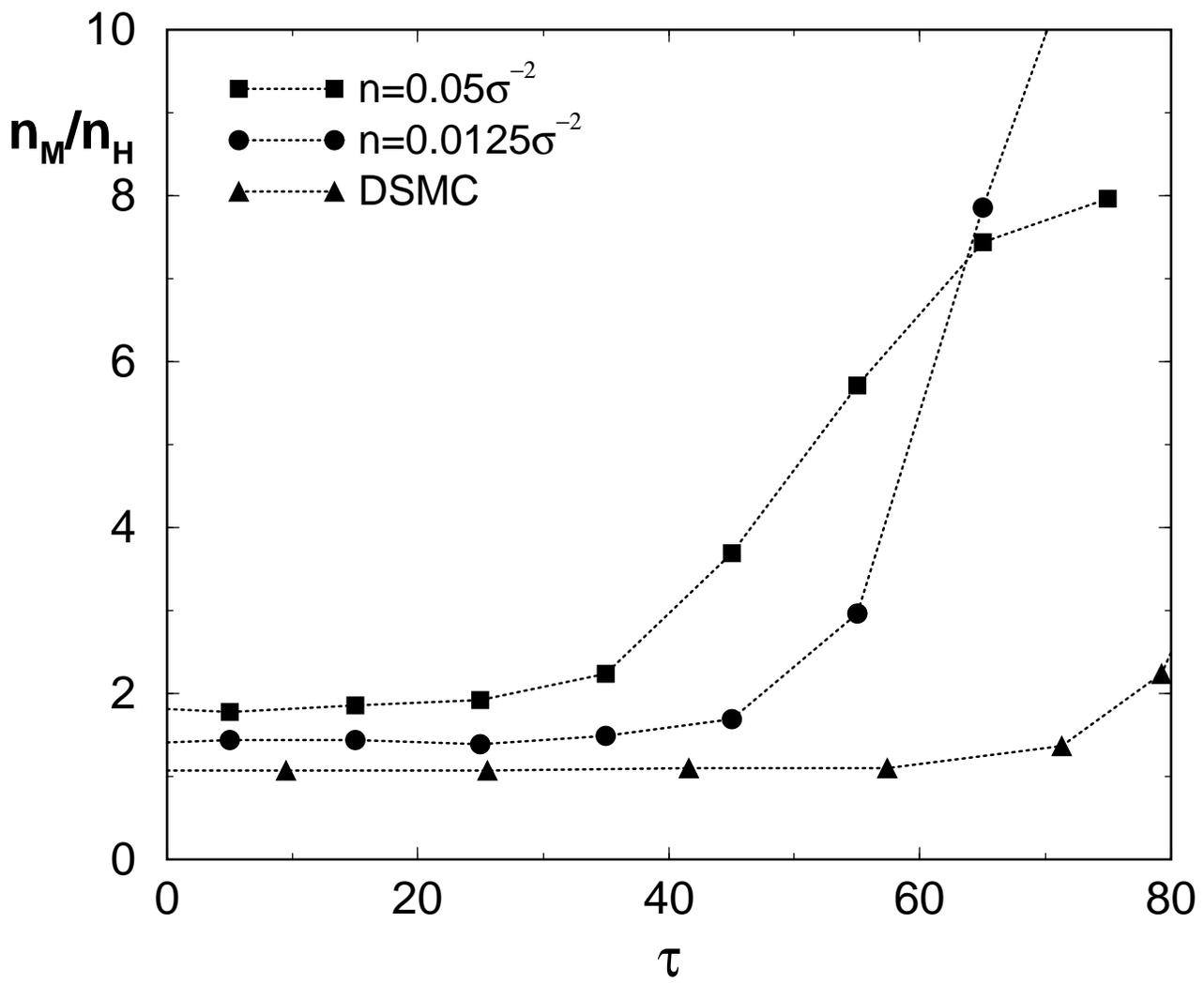
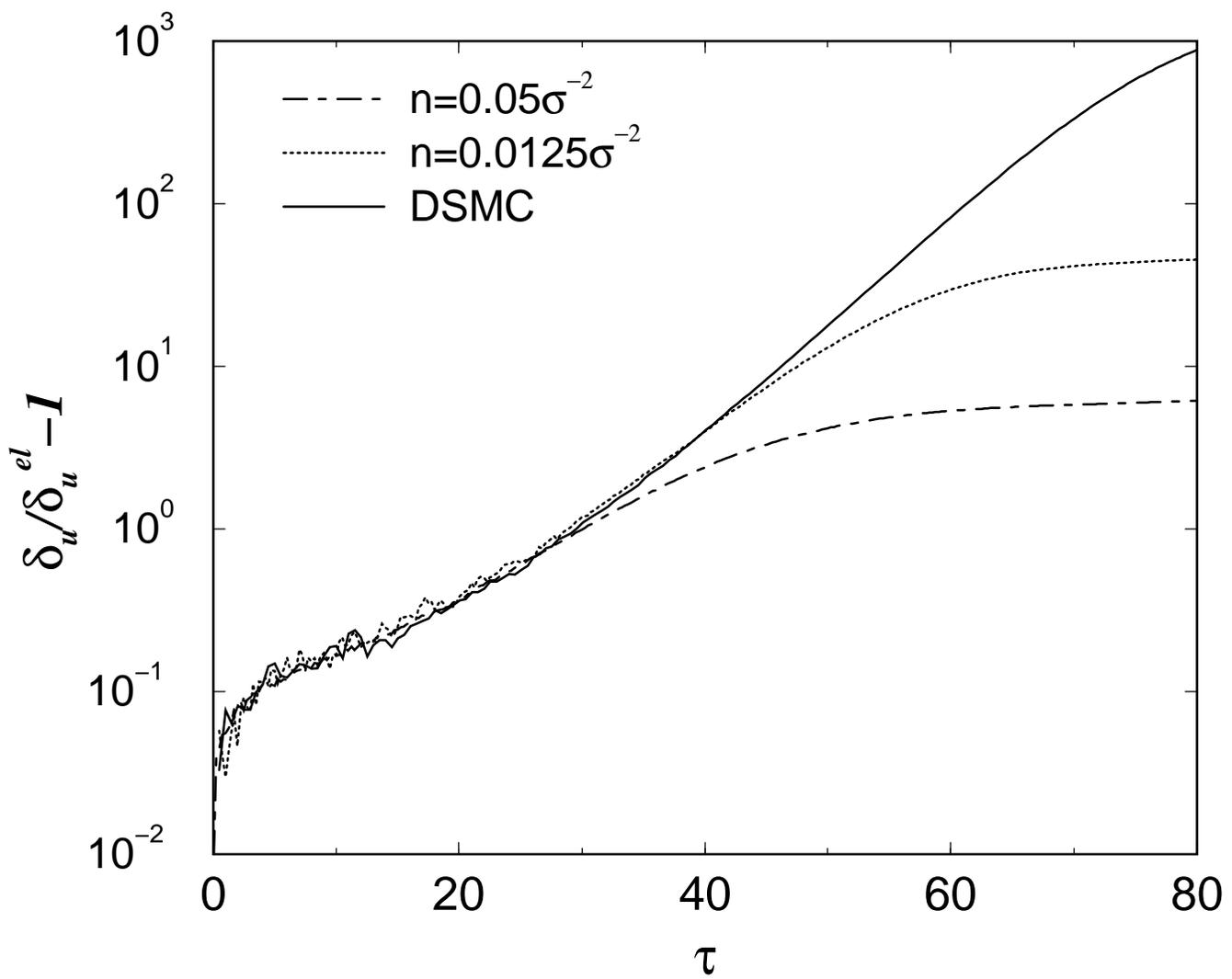
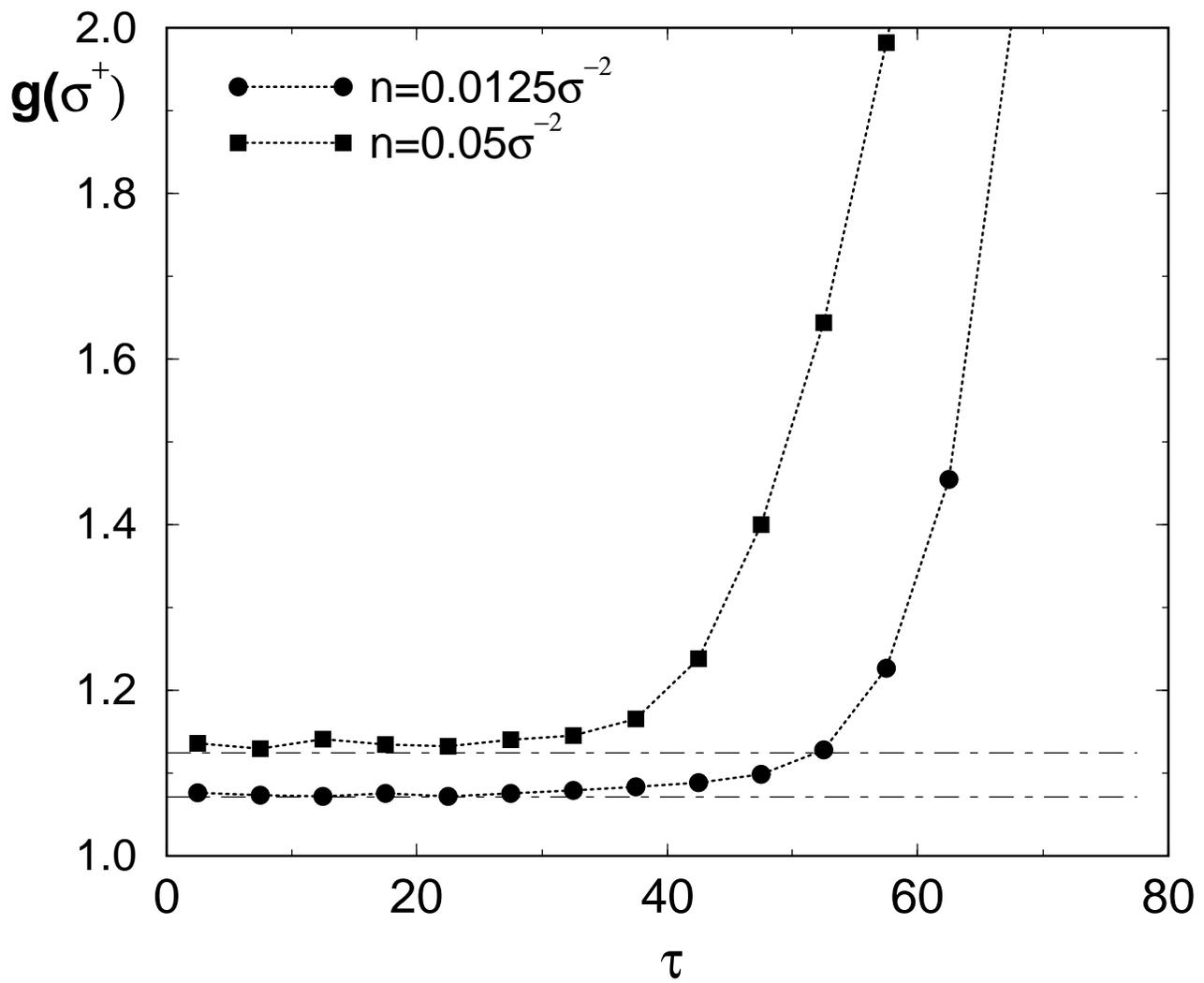
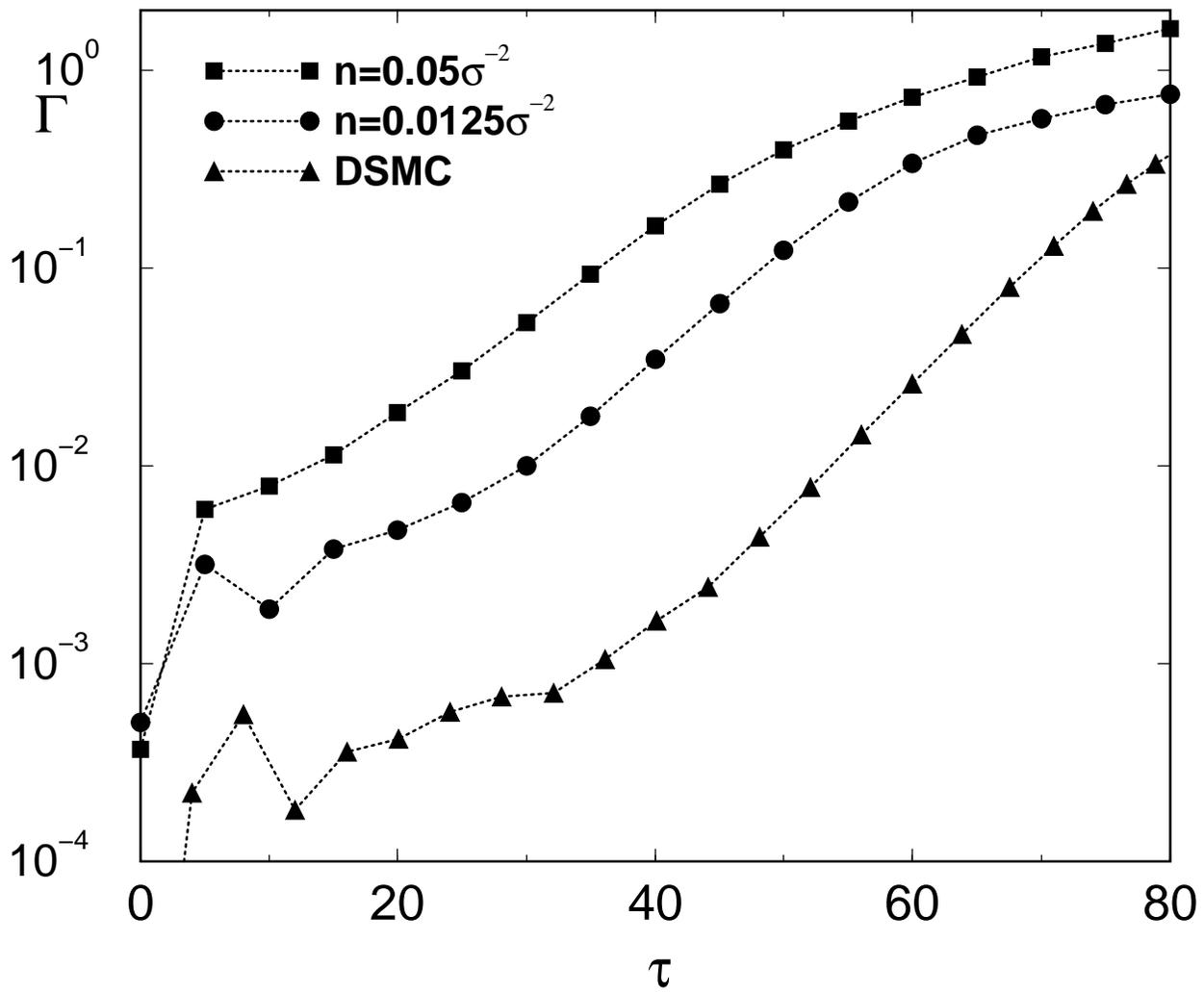
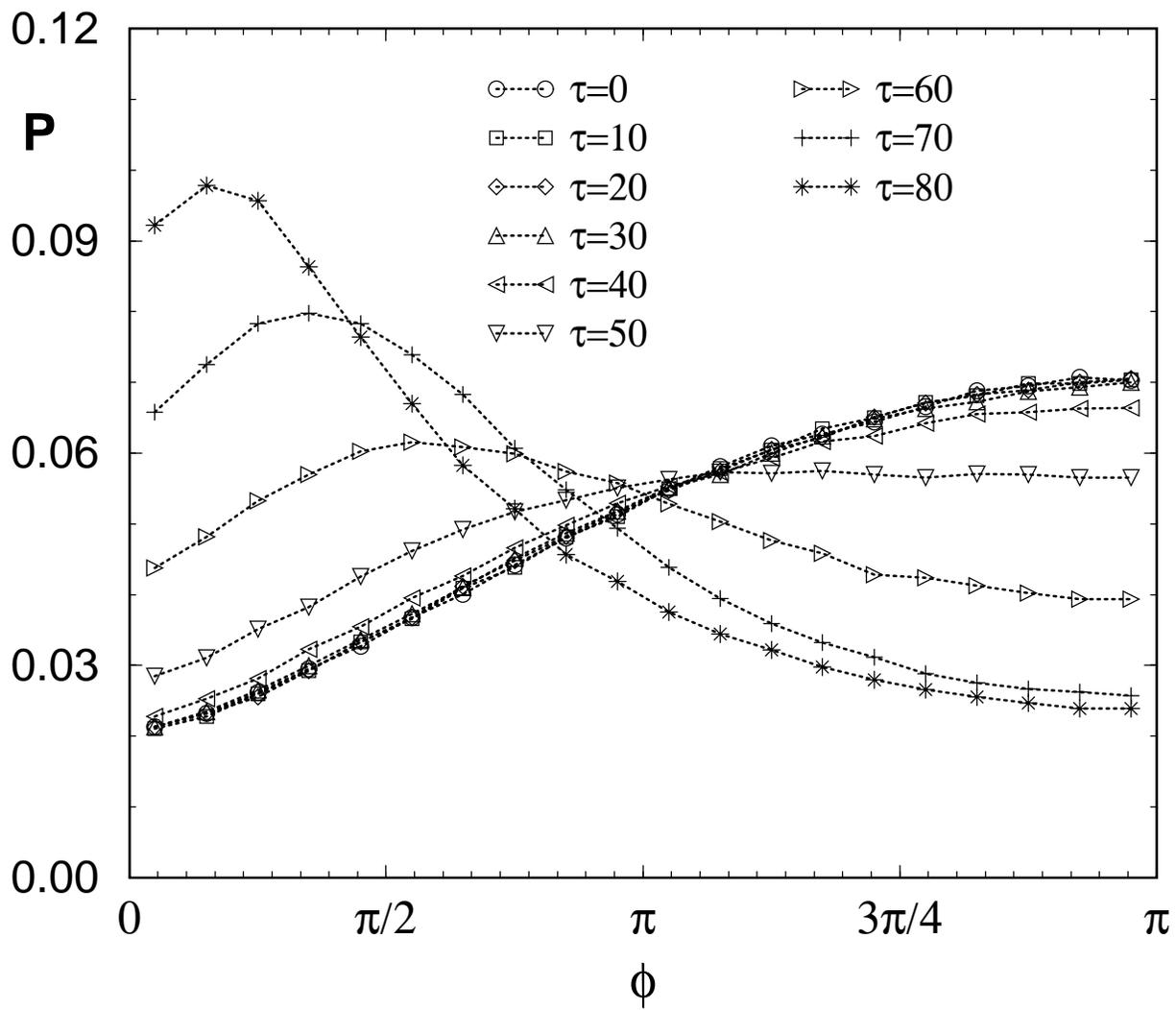
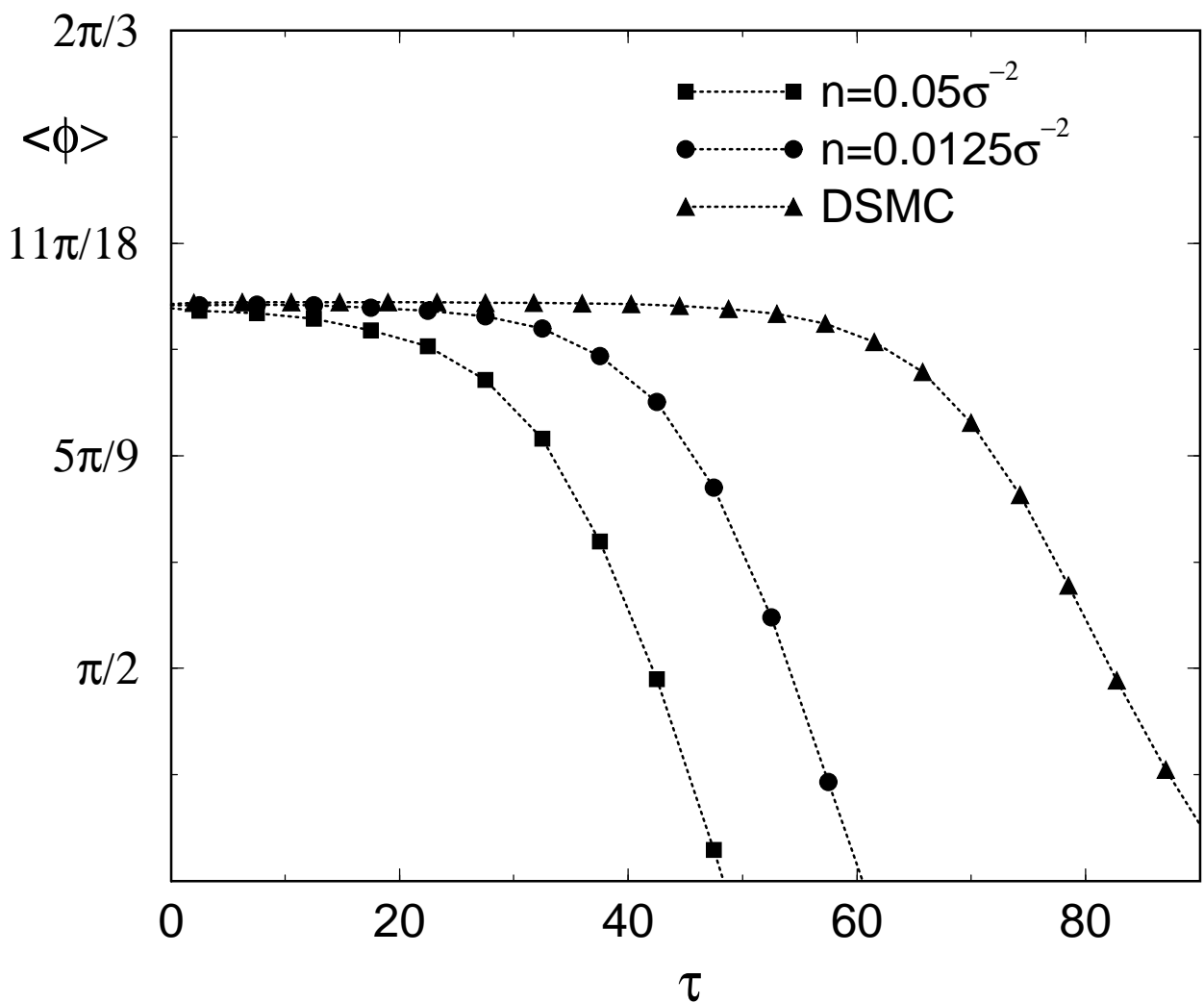
\end{document}